\documentclass{article}

\usepackage[dblblindworkshop, final]{neurips_2026}

\usepackage[utf8]{inputenc}
\usepackage[T1]{fontenc}
\usepackage{hyperref}
\usepackage{url}
\usepackage{booktabs}
\usepackage{amsfonts}
\usepackage{nicefrac}
\usepackage{microtype}
\usepackage{xcolor}
\usepackage{graphicx}
\graphicspath{{./}{figs/}}
\usepackage{threeparttable}

\title{Retention-Constrained Post-Training Quantization of Cellpose--SAM
for Stem Cell Microscopy}

\workshoptitle{LatinX in AI Workshop}

\author{%
  Sebasti\'an A.\ Cruz Romero\thanks{Corresponding author.} \\
  Capic\'u Technologies \\
  \texttt{scruzromero@capicu.ai} \\
}

\begin{document}

\maketitle

\begin{abstract}
Induced pluripotent stem cell (iPSC) culture increasingly relies on segmentation foundation models, yet deployment on laboratory CPUs and edge hardware requires compression schemes that are both efficient and auditable. We present a deployment-oriented evaluation of compressed Cellpose--SAM using a pre-specified retention criterion: the 95\% cluster-bootstrap interval of mean change from FP32 must remain above a fixed $-0.02$ margin for every imaging modality. On a stratified 176-field panel spanning BBBC038 nuclei, BBBC039 U2OS fluorescence, and NIST iPSC images across density regimes, weight-only W8A16 preserves instance F1 across all modalities. A sensitivity-guided mixed W4/W8 scheme, using four INT8 exceptions, achieves a $6.76\times$ reduction in weight storage with no observed catastrophic failures ($0/176$ fields), matching W8A16 at this sample size. In contrast, ternary weight-only quantization achieves $12.08\times$ compression but fails catastrophically on $169/176$ fields. These results demonstrate that compression should be evaluated by modality-stratified downstream retention rather than single-number accuracy, and establish a reproducible protocol for auditing compressed foundation models in regulated stem-cell imaging.
\end{abstract}

\section{Introduction}
\label{sec:intro}

Induced pluripotent stem cells (iPSCs), first reprogrammed from somatic cells almost two decades ago~\citep{takahashi2006ipsc,yu2007ipsc}, are now a foundational substrate for translational biology and for the autologous and allogeneic cell-therapy manufacturing workflows that have emerged in the past decade~\citep{doulgkeroglou2020automation}. Longitudinal microscopy is intrinsic to iPSC culture: density counts, colony-morphology descriptors, and confluence estimates time passaging, guide differentiation protocols, and gate release testing for cell products. Deep learning has become the default approach to these image-analysis steps~\citep{moen2019deep,meijering2020bird,kusumoto2019cnn,waisman2019ipsc}, and the recent generation of segmentation foundation models (Segment--Anything~\citep{kirillov2023sam}, SAM\,2~\citep{ravi2024sam2}, MedSAM~\citep{ma2024medsam}, StarDist~\citep{weigert2020stardist}, and the Cellpose family~\citep{stringer2021cellpose,pachitariu2022cellpose2,stringer2025cellposesam}) generalizes across imaging modalities without the per-assay retraining that dominated prior practice. The generalization is real, and it is one reason these models are being adopted in laboratory and manufacturing workflows even where the evaluation methodology is less mature than in comparable clinical AI settings~\citep{moor2023foundation}.

The inference cost is not free: a $\sim$$305$\,M-parameter foundation model is a heavy load for the CPU and edge-adjacent hardware that characterizes lab benches, incubator-adjacent imaging stations, and small-scale manufacturing floors, where GPU accelerators are frequently absent for reasons of physical placement, validation cost, or regulated environment control. In such settings, compression is not an optimization; it is a deployment requirement. Post-training quantization is the family of methods that most closely matches this posture, because it does not require additional labeled training data and can operate on a released FP32 checkpoint. But choosing among compression schemes cannot be reduced to a single accuracy number. The useful question for a stem-cell imaging workflow is not ``which scheme has the highest AP,'' but ``which schemes are safe to deploy on the imaging modalities I actually run, under a margin I can defend, with an audit trail an independent reviewer can re-derive?''

We propose a pre-specified retention protocol organized around four elements: (i) a criterion defined at a downstream metric (paired instance F1 at IoU $0.5$, together with AP endpoints~\citep{lin2014coco}), not at an output-tensor proxy; (ii) a margin ($-0.02$ mean change from FP32) fixed before any hold-out data are examined; (iii) per-(scheme, imaging modality) verdicts computed with a cluster-bootstrap 95\% interval over experimental units, so a passing verdict on one modality cannot silently obscure a failing verdict on another; and (iv) released audit artifacts so downstream users can re-derive the verdict on their own data.

\section{Related Work}
\label{sec:related}

\paragraph{iPSC biology and manufacturing analytics.} The reprogramming of somatic cells to a pluripotent state~\citep{takahashi2006ipsc,yu2007ipsc} launched a research and manufacturing pipeline that now underlies large-scale cell-therapy production~\citep{doulgkeroglou2020automation}. Automated microscopy is a routine element of iPSC culture and process control; convolutional and deep-learning approaches for stem-cell image analysis were surveyed by~\citet{kusumoto2019cnn}, and phenotype-prediction work at early differentiation was demonstrated by~\citet{waisman2019ipsc}. Colony-scale morphology characterization from large microscopy images has been a NIST focus~\citep{bajcsy2018enabling,chalfoun2016mist}, and the density regime tiles used here are drawn from that dataset.

\paragraph{Segmentation foundation models for cellular imaging.} The Cellpose family introduced dynamics-based flow integration as the decoder for cellular instance segmentation~\citep{stringer2021cellpose,pachitariu2022cellpose2}, and Cellpose--SAM~\citep{stringer2025cellposesam} combines that decoder with a Segment--Anything visual encoder~\citep{kirillov2023sam,ravi2024sam2}. Adjacent biomedical foundation models (MedSAM~\citep{ma2024medsam}, StarDist~\citep{weigert2020stardist}) occupy the same deployment surface but decode via mask tokens or star-convex polyhedra rather than flow integration. Broader cellular deep learning trends are surveyed by~\citet{moen2019deep} and~\citet{meijering2020bird}, and the case for generalist medical AI in this space is made by~\citet{moor2023foundation}.

\paragraph{Post-training quantization for deployment.} Post-training quantization matches the CPU and edge deployment posture that characterizes lab-bench imaging: it does not require additional labeled training data and operates on a released FP32 checkpoint. \citet{jacob2018quantization} formalized integer-arithmetic quantized inference, \citet{krishnamoorthi2018quantizing} and \citet{nagel2021white} extended it into per-layer evaluation practice, and modern PTQ recipes include AdaRound~\citep{nagel2020adaround}, BRECQ~\citep{li2021brecq}, GPTQ~\citep{frantar2022gptq}, LLM.int8()~\citep{dettmers2022llmint8}, and AWQ~\citep{lin2024awq}, with a broader survey in~\citet{gholami2022survey}.

\paragraph{Evaluation methodology and retention criteria.} The medical-image-analysis community has produced explicit guidance on the pitfalls of scalar per-pixel proxies and pooled reporting~\citep{maierhein2018rankings,maierhein2022metrics,reinke2024pitfalls}, and on segmentation-metric sample-size and confidence-interval design~\citep{islam2018clt,varoquaux2022evaluating}. Cluster-bootstrap inference~\citep{efron1994bootstrap,diciccio1996bootstrap} is the standard tool for CI construction over exchangeable experimental units. Pre-registration norms for confirmatory analyses~\citep{nosek2018preregistration} motivate fixing the retention margin before any hold-out data are examined (in our study the margin is pre-specified internally rather than externally registered), and reproducibility reporting in machine learning~\citep{pineau2021reproducibility} informs the release path for the audit artifacts.

\section{Methods}
\label{sec:methods}

\begin{figure*}[t]
  \centering
  \includegraphics[width=\textwidth]{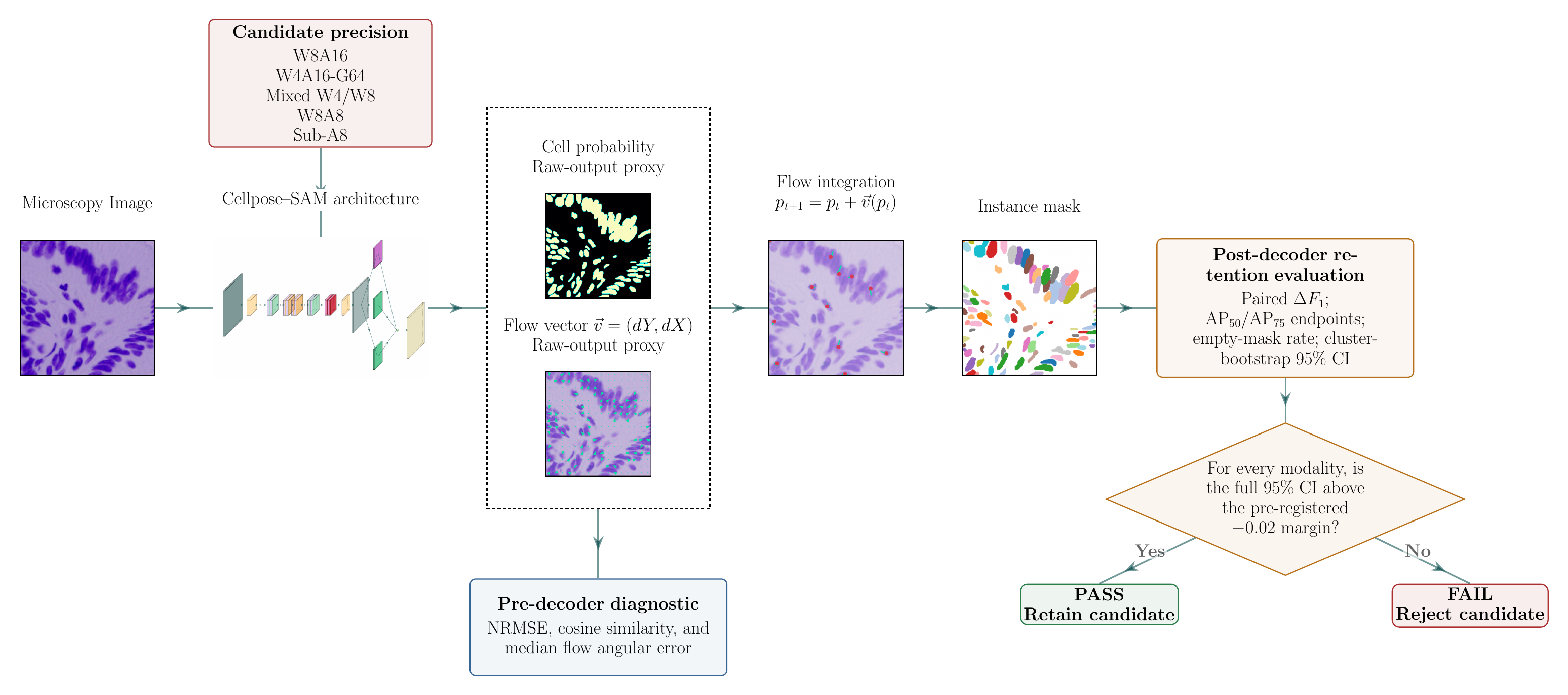}
  \caption{\textbf{Retention protocol at a glance.} Each candidate precision (W8A16, W4A16-G64, mixed W4/W8, W8A8, sub-A8) is
  applied to Cellpose--SAM and traced through the flow-integration decoder to a decoded instance mask. Raw-output taps on the
  cell-probability map and $(dY, dX)$ flow field feed a pre-decoder diagnostic (NRMSE, cosine similarity, median flow angular
  error) that reports numerical distortion but does not gate the decision. The accept/reject verdict is post-decoder: paired
   $\Delta F_1$ at IoU $0.5$, AP$_{50}$/AP$_{75}$ endpoints, empty-mask rate, and a $2{,}000$-draw cluster-bootstrap $95\%$ CI over
   experimental units. A candidate is retained only if the entire interval clears the pre-specified $-0.02$ margin on every
   imaging modality in the panel.}
  \label{fig:retention-protocol}
\end{figure*}

\paragraph{Model.} Cellpose--SAM (checkpoint \texttt{cpsam\_v2}, $\sim$$305$\,M parameters, FP32 storage $1{,}162.07$ MiB), a promptless variant of the Segment--Anything family that produces per-pixel $dY$, $dX$, and cell-probability channels reduced to instance masks by the dynamics-based seeding stage introduced with Cellpose.

\begin{figure*}[t]
\centering
\includegraphics[width=\textwidth]{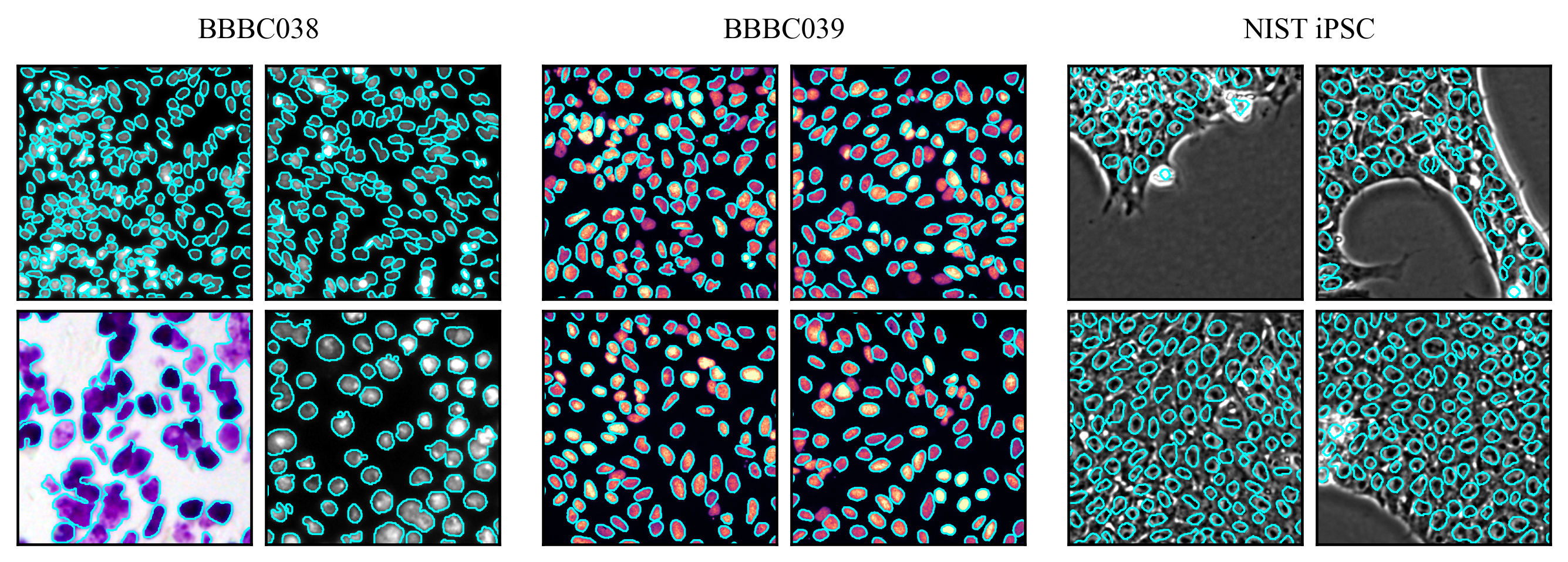}
\caption{\textbf{Retention panel gallery.} Representative fields from the three hold-out modalities used by the retention protocol: BBBC038 heterogeneous nuclei (left, fluorescence / brightfield / phase-contrast), BBBC039 U2OS nuclei fluorescence (middle), and NIST iPSC phase-contrast tiles spanning low, medium, and high culture density (right). Cyan overlays are supplied reference-instance contours. The three modalities differ in cell type, illumination, and object density: BBBC038 mixes staining protocols and can contain hundreds of small nuclei per field, BBBC039 is a homogeneous fluorescence dataset with well-separated ellipsoidal nuclei, and NIST iPSC covers colony morphology from sparse post-passage plating through pre-passage confluence. The retention criterion in Section~\ref{sec:methods} passes only when the entire cluster-bootstrap CI on paired instance F1 clears the pre-specified $-0.02$ margin on every modality shown here.}
\label{fig:methods-panel-gallery}
\end{figure*}

\paragraph{Panel.} A stratified public panel of $176$ hold-out fields covering three imaging modalities: BBBC038 ($n=70$, nuclei across fluorescence, brightfield, phase contrast), BBBC039 ($n=34$, U2OS nuclei fluorescence), and NIST iPSC ($n=72$ across low, medium, and high culture density regimes, aggregated over $3$ source-image experimental units). Sensitivity, bit allocation, and calibration use a disjoint development split. Cluster-bootstrap CIs use the experimental unit ($107$ across the retention panel) as the resampling atom, so BBBC field-level fields resample individually while NIST tiles resample by their source image. Sample-size design targets $\sim$$4\%$ CI widths on segmentation metrics under the heuristics of~\citet{varoquaux2022evaluating} and~\citet{islam2018clt}. Figure~\ref{fig:methods-panel-gallery} shows one representative field per modality with cyan reference-instance contours.

\paragraph{Retention criterion.} For each (scheme, modality) pair, let $\Delta$ be the mean change from FP32 in the paired downstream metric across hold-out fields; we form a $2{,}000$-draw cluster-bootstrap 95\% interval for $\mathbb{E}[\Delta]$ over experimental units. A scheme meets the criterion on that modality if the entire interval lies above the pre-specified $-0.02$ margin. In addition, we report a per-scheme catastrophic-failure rate (the fraction of hold-out fields on which mask output falls below an absolute quality floor) with a bootstrap 95\% interval over experimental units. The margin was fixed before any hold-out evaluation.

\paragraph{Compression schemes.} Weight-only \mbox{W8A16}, \mbox{W4A16-G64} (group size $64$), and ternary \mbox{W2A16-G64}; a calibrated \mbox{W8A16-QDQ-obs} scheme (weight-only packing with per-tensor activation QDQ instrumentation; packed weight storage identical to W8A16, see Table~\ref{tab:storage}), distinct from the whole-graph W8A8 configurations that are known to collapse on this decoder and that are shown for reference in Figure~\ref{fig:mechanism}D; and a sensitivity-guided mixed scheme whose four highest-sensitivity operators (identified by a one-operator-at-a-time W4 perturbation over $100$ eligible Linear/Conv2d operators on the development split) are retained at INT8 while the remainder use W4.

\section{Results}
\label{sec:results}

Table~\ref{tab:retention} reports per-(scheme, modality) retention numbers on the $176$-field hold-out. Weight-only W8A16 preserves paired instance F1 tightly across all three modalities ($\Delta F_1 \in \{+0.0004, +0.0003, +0.0006\}$, all CIs contained in the $[-0.001, +0.002]$ band). Weight-only W4A16-G64 stays within the margin on all three modalities ($\Delta F_1 \in \{-0.0000, +0.0006, -0.0027\}$), with the widest arm ($-0.0027$ on NIST iPSC) still comfortably above the $-0.02$ threshold. Calibrated W8A16-QDQ-obs (weight-only packing) preserves F1 within the margin on every modality ($\Delta F_1 \in \{-0.0028, +0.0003, +0.0004\}$). The sensitivity-guided mixed W4/W8 scheme passes on all three modalities as well ($\Delta F_1 \in \{+0.0017, +0.0004, -0.0012\}$). Ternary W2A16-G64 fails the criterion catastrophically on the segmentation-dominant modalities, with $\Delta F_1 = -0.841$ $[-0.882, -0.793]$ on BBBC038 and $-0.907$ $[-0.929, -0.876]$ on BBBC039; the NIST iPSC arm is floor-limited on FP32 so the paired F1 drop is only $-0.023$ $[-0.054, -0.002]$, but the compressed model produces essentially no correct instances there either.

\begin{table}[t]
\centering
\small
\begin{threeparttable}
\caption{Paired instance-$F_1$ change (mean $\Delta F_1$ [95\% CI]) from FP32 at IoU $=0.5$ across
imaging modalities. Values are $\Delta F_1$ with 95\% cluster-bootstrap
confidence intervals of 2,000 draws. Sample sizes are BBBC038 ($n=70$),
BBBC039 ($n=34$), and NIST iPSC ($n=72$).}
\label{tab:retention}
\setlength{\tabcolsep}{4pt}
\renewcommand{\arraystretch}{1.15}
\begin{tabular*}{\linewidth}{@{\extracolsep{\fill}}lccc@{}}
\toprule
& \textbf{BBBC038} & \textbf{BBBC039} & \textbf{NIST iPSC} \\
\textbf{Scheme} & $n=70$ & $n=34$ & $n=72$ \\
\midrule
W8A16 &
\shortstack[r]{$+0.0004$\\{\footnotesize[$-0.0005,\,+0.0016$]}} &
\shortstack[r]{$+0.0003$\\{\footnotesize[$\phantom{-}0.0000,\,+0.0007$]}} &
\shortstack[r]{$+0.0006$\\{\footnotesize[$-0.0001,\,+0.0017$]}} \\
W4A16-G64 &
\shortstack[r]{$-0.0000$\\{\footnotesize[$-0.0052,\,+0.0051$]}} &
\shortstack[r]{$+0.0006$\\{\footnotesize[$-0.0012,\,+0.0033$]}} &
\shortstack[r]{$-0.0027$\\{\footnotesize[$-0.0047,\,-0.0012$]}} \\
W8A16-QDQ-obs &
\shortstack[r]{$-0.0028$\\{\footnotesize[$-0.0070,\,+0.0007$]}} &
\shortstack[r]{$+0.0003$\\{\footnotesize[$-0.0014,\,+0.0022$]}} &
\shortstack[r]{$+0.0004$\\{\footnotesize[$-0.0029,\,+0.0063$]}} \\
MW4-SW8 &
\shortstack[r]{$+0.0017$\\{\footnotesize[$-0.0027,\,+0.0055$]}} &
\shortstack[r]{$+0.0004$\\{\footnotesize[$-0.0008,\,+0.0018$]}} &
\shortstack[r]{$-0.0012$\\{\footnotesize[$-0.0025,\,+0.0010$]}} \\
\midrule
W2A16-G64 &
\shortstack[r]{$-0.841$\\{\footnotesize[$-0.882,\,-0.793$]}} &
\shortstack[r]{$-0.907$\\{\footnotesize[$-0.929,\,-0.876$]}} &
\shortstack[r]{$-0.023$\\{\footnotesize[$-0.054,\,-0.002$]}} \\
\bottomrule
\end{tabular*}
\begin{tablenotes}[flushleft]
\footnotesize
\item $\Delta F_1$ denotes the paired difference relative to FP32.
Intervals are percentile cluster-bootstrap 95\% CIs over experimental units.
\end{tablenotes}
\end{threeparttable}
\end{table}

The panel-level catastrophic-failure rate in Table~\ref{tab:catastrophic} concentrates the same picture into a single interval per scheme. W8A16 and the sensitivity-guided mixed W4/W8 scheme deliver $0/176$ observed catastrophic fields; because the percentile bootstrap cannot generate a nonzero rate when no unit fails, we quote a rule-of-three upper bound of $3/107 \approx 0.028$ over experimental units in place of the degenerate $[0,0]$ interval, so the strongest claim these two schemes support is ``no failure observed at this sample size,'' not ``failure rate is zero.'' W8A16-QDQ-obs and W4A16-G64 have $1/176$ catastrophic fields each with cluster-bootstrap interval $[0.000, 0.001]$; ternary W2A16-G64 has $169/176$ catastrophic fields with interval $[0.949, 1.000]$. This is the compact form of the deployment answer for iPSC monitoring: four schemes are safe candidates at margin, and one is unusable at any compression advantage.

\begin{table}[t]
\centering
\small
\caption{Catastrophic-failure rate on the 176-field hold-out panel
(107 experimental units). Rates are estimated at the experimental-unit
level with 2,000-draw cluster-bootstrap 95\% percentile intervals.
Zero-event schemes use the rule-of-three upper bound $3/107 \approx 0.028$.
A catastrophic failure denotes a field whose compressed output falls
below the pre-specified absolute quality floor.}
\label{tab:catastrophic}
\begin{tabular}{lrr}
\toprule
Scheme & Failed / Total & Rate [95\% CI] \\
\midrule
W8A16                 & 0 / 176   & 0.000 [0.000, 0.028]$^{\dagger}$ \\
mixed-W4-sensitive-W8 & 0 / 176   & 0.000 [0.000, 0.028]$^{\dagger}$ \\
W8A16-QDQ-obs         & 1 / 176   & 0.001 [0.000, 0.001] \\
W4A16-G64             & 1 / 176   & 0.001 [0.000, 0.001] \\
W2A16-G64             & 169 / 176 & 0.979 [0.949, 1.000] \\
\bottomrule
\end{tabular}

\vspace{2pt}
\begin{minipage}{0.94\linewidth}
\footnotesize
$^{\dagger}$ No failures were observed; the percentile bootstrap is
degenerate at zero, so the reported upper bound is the rule-of-three
estimate $3/107$ over experimental units.
\end{minipage}
\end{table}

Table~\ref{tab:storage} pairs the retention outcome with the compression profile that a lab-bench deployment cares about. W8A16 reduces weight storage $3.93\times$; W4A16-G64, $6.90\times$; the sensitivity-guided mixed W4/W8 scheme, $6.76\times$; and ternary W2A16-G64, $12.08\times$. The mixed scheme therefore combines the compression profile of the W4-based schemes ($6.76\times$) with a catastrophic-failure count tied for tightest with W8A16 on our panel ($0/176$ observed, rule-of-three upper bound $0.028$). In stem-cell terms, the NIST iPSC arm covers three culture states that are traversed in every real iPSC workflow (sparse plating early after passage, mid-log growth, and pre-passage confluence when overlap and touching become dominant), and per-modality retention passing on that arm is a necessary but not sufficient condition for iPSC deployment. Necessary, because iPSC monitoring protocols sample the full density trajectory and a scheme that silently degrades in the high-density regime would delay passaging decisions or contaminate downstream release-testing counts. Not sufficient, because FP32 itself does not segment reliably on high-density iPSC in this panel (Table~\ref{tab:retention}, and see the AP@0.75/0.90 floor-limitation reported below); the sufficient condition is a NIST-analogue arm where the FP32 reference actually detects instances, which would need to be assembled from a workflow-specific dataset before deployment. On the present panel the NIST verdicts primarily demonstrate that the protocol correctly flags a floor-limited modality rather than that the compressed model preserves detection quality on iPSC.

\begin{table}[t]
\centering
\small
\begin{threeparttable}
\caption{Compression profile for the evaluated schemes. Weight-storage ratios are computed on quantized weight blobs; FP32 scale and reconstruction metadata retained on disk for auditable inference are excluded from the ratio.}
\label{tab:storage}
\begin{tabular}{lccr}
\toprule
Scheme & Storage (MiB) & Compression & Notes \\
\midrule
FP32 (reference)      & 1162.07 & 1.00$\times$  & Baseline \\
W8A16                 & 295.78  & 3.93$\times$  & Tightest retention \\
W8A16-QDQ-obs (weight-only)& 295.78  & 3.93$\times$  & Tightest retention \\
W4A16-G64             & 168.47  & 6.90$\times$  & Marginal retention on NIST \\
MW4-SW8               & 171.86  & 6.76$\times$  & INT8 exemptions\tnote{a} \\
W2A16-G64             & 96.21   & 12.08$\times$ & Fails retention \\
\bottomrule
\end{tabular}
\begin{tablenotes}[flushleft]
\footnotesize
\item[a] INT8 exemptions: \texttt{blocks.0.mlp.lin2}, \texttt{blocks.23.attn.qkv}, \texttt{neck.2}, and \texttt{out}.
\end{tablenotes}
\end{threeparttable}
\end{table}

\begin{figure*}[t]
\centering
\includegraphics[width=.9\textwidth]{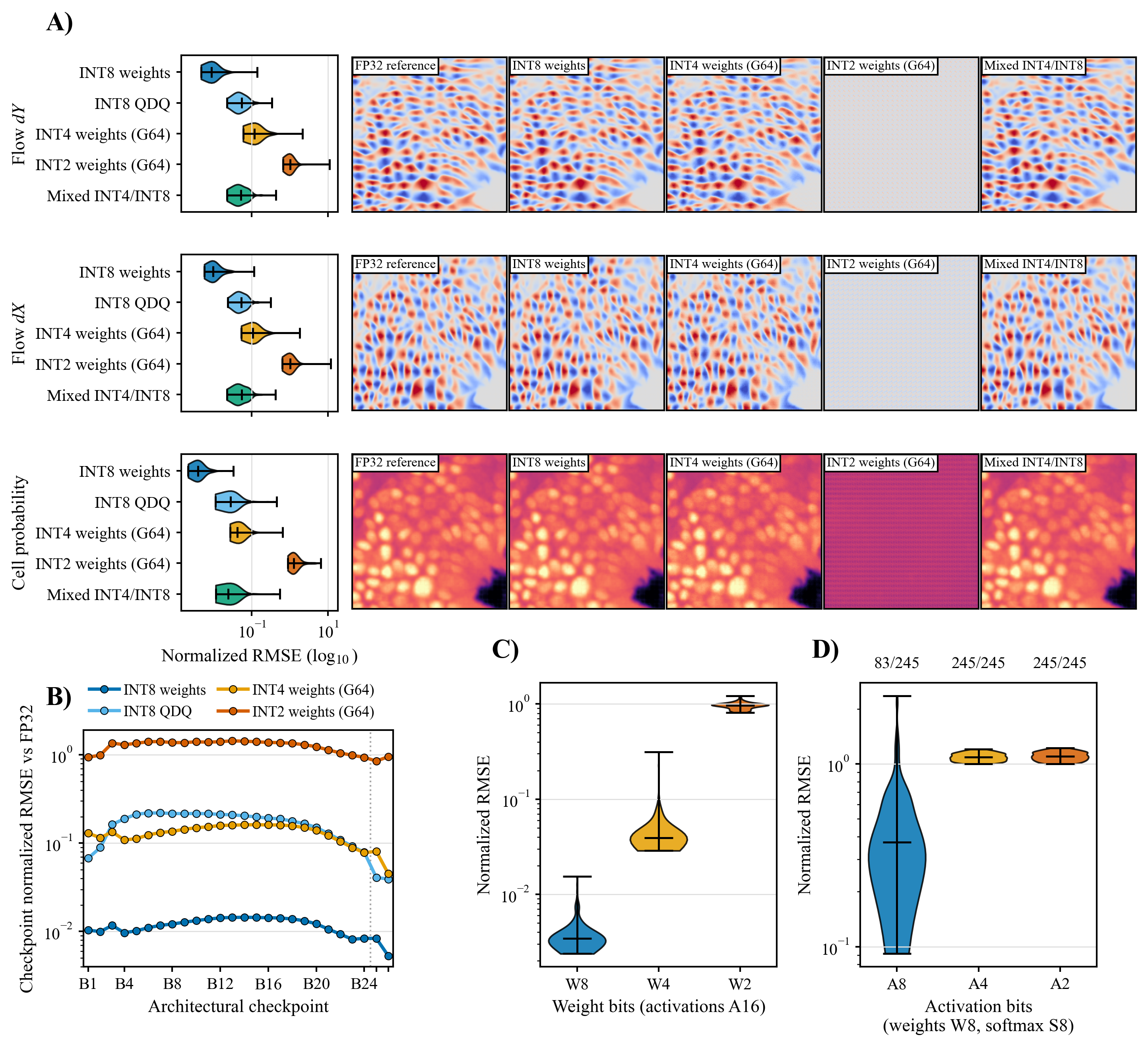}
\caption{\textbf{Mechanism panel supporting the retention verdicts.} \textbf{A)} per-field raw-output NRMSE against FP32 (log-scale violins per scheme; rows: Flow $dY$, Flow $dX$, cell probability) with a qualitative BBBC038 composite; ternary INT2 detaches by $\sim$$10\times$ in NRMSE, matching the $169/176$ catastrophic fraction in Table~\ref{tab:catastrophic}. \textbf{B)} checkpoint-level NRMSE through the $26$ architectural checkpoints (B1--B26; ViT-L transformer blocks plus patch-embedding and neck taps): weight-only damage localizes rather than accumulates. \textbf{C)} weight-bit sweep at fixed A16: median NRMSE moves smoothly with bit width. \textbf{D)} activation-bit sweep at fixed W8/S8, empty-mask counts on the broader $245$-field whole-graph panel (BBBC038 $n=99$, BBBC039 $n=47$, NIST iPSC $n=99$; distinct from the $176$-field retention holdout): A8 retains most fields ($83/245$ empty), while A4 and A2 collapse to $245/245$, evidence that sub-A8 rejection is a categorical decoder failure rather than a threshold artifact.}
\label{fig:mechanism}
\end{figure*}

Figure~\ref{fig:mechanism} characterizes why the retention verdicts in Tables~\ref{tab:retention}--\ref{tab:catastrophic} land where they do. The four panels are diagnostic rather than dispositive (the deployment decision remains with the pre-specified mask-level margin), but they show that each verdict has a physical footprint at the tensor and encoder-block level, which is what an auditable release process needs when the reader is not the person who ran the pipeline.

\section{Discussion \& Conclusion}
\label{sec:discussion}

Compression choices for a deployed foundation model are not a single-scalar optimization problem. The useful answer in a stem-cell imaging setting is a verdict on which schemes are safe to deploy on which imaging modalities, computed against a pre-specified margin, with enough audit trail to be re-derived by an independent reviewer on independent data. Under the retention protocol proposed here, weight-only W8A16 and the sensitivity-guided mixed W4/W8 scheme tie at $0/176$ observed catastrophic fields (rule-of-three upper bound $0.028$ over $107$ experimental units); the mixed scheme deepens the compression from $3.93\times$ to $6.76\times$ at no measurable loss on our panel. Calibrated W8A16-QDQ-obs (weight-only packing) and W4A16-G64 sit one failed field back at $[0.00, 0.001]$; and ternary weight-only quantization at $12.08\times$ is unusable at the margin regardless of the storage advantage ($169/176$ failed fields, interval $[0.95, 1.00]$). The evaluation protocol itself is portable across cell-imaging deployments: it requires a retention target appropriate to the downstream use, a stratified panel that covers the distributional axes that matter for the deployment, an experimental unit over which resampling makes physical sense, and a release path for the audit artifacts. The specific bit widths and operator selections that pass are properties of Cellpose--SAM and of the mask-level retention target, and should be re-derived, not inherited, by users deploying different decoders, modalities, or margins.

\section{Limitations}
\label{sec:limits}

The retention verdicts scope to the panel modalities evaluated here; a passing verdict on BBBC038, BBBC039, or NIST iPSC does not certify the same scheme on modalities with different statistics (3D volumetric microscopy, time-lapse phase contrast at extended intervals, or non-nuclear stains). The $-0.02$ margin is appropriate to a monitoring pipeline that can absorb a bounded per-modality F1 drop; regulatory release-testing pipelines require a tighter margin, and the protocol scales but the verdicts do not transfer. The specific passing bit widths and operator selection are properties of Cellpose--SAM's promptless flow-integration decoder and will not directly transfer to mask-token or U-Net decoders; the protocol is the transferable element.

\bibliographystyle{plainnat}
\bibliography{lxai_wrksp_cruzromero/ref}

\end{document}